\documentclass[prd,tightenlines,superscriptaddress]{revtex4}
\usepackage{enumerate}
\usepackage{amssymb}
\usepackage{amsfonts}
\usepackage{mathrsfs}
\usepackage{latexsym}
\usepackage{bm}
\usepackage{amscd}
\usepackage{graphicx}
\usepackage{epsfig}
\usepackage{hhline,multirow}
\usepackage{dcolumn}
\usepackage{url}
\usepackage[colorlinks=true,linkcolor=blue]{hyperref}
\usepackage{color}
\usepackage{indentfirst}
\usepackage{subfigure}
\usepackage{psfrag}
\usepackage{slashed}
\usepackage{float}
\usepackage{setspace}

\begin{document}

\title{Color confinement, dark matter and the missing of anti-matter}
\author{P. Wang}
\affiliation{Institute of High Energy Physics, CAS, Beijing 100049, China}

\begin{abstract}

QCD is the fundamental theory to describe the strong interaction, where quarks and gluons have the color degrees of freedom. However, a single quark or gluon can not be separated out and all observable particles are color singlet states. Color confinement or quark confinement conjecture can be proved by considering not only the strong interaction but also the electroweak interaction which is $SU(3)_c$ invariant. Any measurable state has to be color singlet is the direct consequence of the common symmetry of the standard model. Color non-singlet objects are created from the big bang when the interaction breaks $SU(3)_c$ symmetry based on the nonlocal Lagrangian. There is nearly no interaction between colored objects and color singlet universe when the momentum transfer is not large enough. Colored objects are reasonable candidates of dark matter and the missing of anti-matter in the universe can also be easily explained. Dark matter can be produced in the laboratory which can be tested by measuring the energy loss and baryon number change in the extremely high energy collisions of particles and anti-particles.

\end{abstract}


\maketitle

The standard model has been well established for many years. Quarks and leptons are fundamental fermions which interact with the gauge fields through the gauge invariant Lagrangian. Different from leptons, quarks have the color degrees of freedom as well as the flavor. The Lagrangian density of quantum chromodynamics (QCD) for the strong interaction is written as 
\begin{equation} \label{eq:LQCD}
{\cal L}^{\rm QCD} = \bar{q}^\alpha_f (i \slashed{D}_{\alpha\beta} - m_f \delta_{\alpha\beta}) q^\beta_f 
- \frac14 {\cal G}^a_{\mu\nu}{\cal G}_a^{\mu\nu},
\end{equation}
where $f$ ($f=u,d,s,c,b,t$) and $\alpha$, $\beta$ ($\alpha, \beta = r,g,b$) are the flavor and color indices 
and $\slashed{D}_{\alpha\beta} = (\delta_{\alpha\beta}\partial_\mu - ig_s \frac{\lambda^a_{\alpha\beta}}{2} {\cal A}_\mu^a) \gamma^\mu$. The Einstein sum rule for repeated indices is used throughout this paper. $\lambda^a (a=1,...,8)$ are the eight Gell-Mann matrices. ${\cal G}^a_{\mu\nu}$ is expressed as
\begin{equation}
{\cal G}^a_{\mu\nu} = \partial_\mu {\cal A}_\nu^a - \partial_\nu {\cal A}_\mu^a 
+ g_s f_{abc} {\cal A}_\mu^b {\cal A}_\nu^c,
\end{equation}
where $f_{abc}$ are the structure constants of $SU(3)$ group.

Though QCD is very successful, especially in the perturbative region \cite{Hooft,Gross,Politzer} and the color degrees of freedom have been confirmed by the experimental measurements of the cross section ratio of hadron production to lepton production from $e^+ e^-$ collisions, there exists
long-standing difficulty of explaining or proving color confinement. In fact, in any $e^+ e^-$ collisions, heavy-ion collisions or other physical processes, we never found colored states. All the final states are color singlet hadrons and colorless leptons and gauge bosons. One may wonder whether color confinement is the consequence of QCD or there is another mechanism for color confinement in addition to the standard model. 

Color confinement is the most amazing phenomenon in QCD and it is quite different from our previous knowledge. Before quark degrees of freedom were discovered, we thought any object can be separated as long as the energy is large enough. The color singlet conjecture was proposed to explain the phenomenon. Since there is no flavor confinement and charge confinement, for many years, the efforts to prove color confinement conjecture have been focused on the special characteristics of the strong interaction. For example, the non-Abelian property of the strong interaction, linear potential of colored quarks simulated on lattice, vacuum structure of QCD, etc, are often applied to explain color confinement \cite{Konishi,Takahashi,Nakamura,Kharzeev,Kuvshinov}. The proof of color confinement in this direction is very difficult because of the nonperturbative behavior of QCD. Up to now, color confinement or color singlet has not been proved theoretically and it remains to be a conjecture explaining the experimental phenomenon.

We will show the $SU(3)_c$ invariance of the electroweak interaction is crucial to understand color confinement which was thought to be a special property of the strong interaction. The key point is that, to explain color confinement of the strong interaction, we need pay our attention on the common symmetry of all the fundamental interactions in the standard model rather than focus only on the strong interaction. Such solution of color confinement is also related to the other two important issues in cosmology and particle physics, dark matter and the missing of anti-matter which are the most unexpected observations about the universe. Both of them remain mysteries and inspire the imagination of model builders. For dark matter, various candidates have been proposed, such as weakly interacting massive particles, axions, sterile neutrinos, primordial black holes, etc, with masses ranging from $10^{-5}$ eV ($10^{-71}$ solar mass) to $10^4$ solar mass \cite{Jungman,Byrne,Peccei,Abe,Dodelson,Carr,Arun}. There are several experiments to detect dark matter particles running for many years that have yielded no positive results so far \cite{Bernabei,Akeribet,Tan,Amole,Agnese,Abdelhameed}. For the large asymmetry between matter and anti-matter in the universe, though $CP$ violation could lead to the asymmetry, this effect predicted in the standard model is too small to account for the absence of anti-matter and motivates the studies of $CP$ violation beyond the standard model \cite{Langacher,Bernreuther}. 

In this paper, we try to understand and prove color confinement from the common symmetry of the standard model. One can verify that the Lagrangian (\ref{eq:LQCD}) is invariant under the local $SU(3)_c$ transformation
\begin{eqnarray} \label{eq:SU3}
\left(\begin{array}{c} q^r_f \\ q^g_f \\ q^b_f \end{array} \right) \rightarrow U(x) 
\left(\begin{array}{c} q^r_f \\ q^g_f \\ q^b_f \end{array} \right), ~~~ 
{\cal A}_\mu^a \frac{\lambda^a}{2} \rightarrow 
U(x) {\cal A}_\mu^a \frac{\lambda^a}{2} U^{-1}(x) + \frac{i}{g_s}U(x)\partial_\mu U^{-1}(x),
\end{eqnarray}
where 
$U(x)={\rm exp}\left( i\theta^a(x) \frac{\lambda^a}{2}\right)$ acting on the color indices. In addition to the local symmetry, the free and interacting Lagrangian of QCD, ${\cal L}_0^{\rm QCD}$ and ${\cal L}_I^{\rm QCD}$, are both globally $SU(3)_c$ invariant. For example, the free Lagrangian is expressed as 
\begin{equation} \label{eq:LQCDfree}
{\cal L}_0^{\rm QCD} = \bar{q}^\alpha_f (i \slashed{\partial} - m_f) q^\alpha_f 
- \frac14 F^a_{\mu\nu}F_a^{\mu\nu},
\end{equation}
where 
\begin{equation}
F^a_{\mu\nu} = \partial_\mu {\cal A}_\nu^a - \partial_\nu {\cal A}_\mu^a .
\end{equation}
Under the global $SU(3)_c$ transformation, the free Lagrangian is invariant as 
\begin{equation} \label{eq:LQCDfree}
{\cal L}_0^{\rm QCD} \rightarrow \bar{q}_f U^{-1} (i \slashed{\partial} - m_f) U q_f  
- \frac12 {\rm Tr} [U F_{\mu\nu} U^{-1} U F^{\mu\nu}U^{-1}] = \bar{q}_f (i \slashed{\partial} - m_f) q_f - \frac12 {\rm Tr} [U^{-1}U F_{\mu\nu} F^{\mu\nu}] 
= {\cal L}_0^{\rm QCD},
\end{equation}
where $U$ is $x$ independent and $F_{\mu\nu} = F_{\mu\nu}^a \lambda_a$. ${\rm Tr}[\lambda_a \lambda_b]= 2\delta_{ab}$ has been used in the above equation.

For the strong interaction, if we neglect the mass difference between $u$ and $d$ quark, QCD is also invariant under the global $SU(2)_f$ transformation 
\begin{eqnarray} \label{eq:SU2}
\left(\begin{array}{c} q^\alpha_u \\ q^\alpha_d \end{array} \right) \rightarrow {\rm exp}\left( i\theta^a \frac{\tau^a}{2}\right)
\left(\begin{array}{c} q^\alpha_u \\ q^\alpha_d \end{array} \right), ~~~ 
{\cal A}_\mu^a \rightarrow {\cal A}_\mu^a ,
\end{eqnarray}
where $\theta^a$ are constants and $\tau^a (a=1,2,3)$ are the three Pauli matrices. This $SU(2)_f$ symmetry is broken for the electroweak interaction because of the mixing of the weak and electromagnetic 
interactions. The Lagrangian density for the electroweak interaction is locally $SU(2)_L \times U(1)_Y$ invariant \cite{Weinberg,Glashow,Salam} and it is written as
\begin{eqnarray}
{\cal L}_I^{\rm EW} = -\frac{g}{\sqrt{2}} \left( J_\mu^+ W^{+\mu} + J_\mu^- W^{-\mu} \right) - \frac{g}{{\rm cos}\theta_w} \left( J^3_{\mu} - {\rm sin}^2\theta_w J_\mu^{\rm em}\right) Z^\mu - e J_\mu^{\rm em}A^\mu,
\end{eqnarray}
where $\theta_w$ is the weak mixing angle. The charged, neutral and electromagnetic currents are expressed as
\begin{eqnarray} \label{eq:currents}
J^{\pm}_\mu &=& \frac12 \bar{q}^{L,\alpha}_i \gamma_\mu (\tau_1 \pm i \tau_2)^{ij} q^{L,\alpha}_j, \nonumber \\
J^3_\mu &=& \frac12 \bar{q}^{L,\alpha}_i \gamma_\mu \tau_3^{ij} q^{L,\alpha}_j, \nonumber \\
J^{\rm em}_\mu &=& J^3_\mu + J^Y_\mu = \frac12 \bar{q}^{L,\alpha}_i \gamma_\mu \tau_3^{ij} q^{L,\alpha}_j 
+  \frac12 Y \bar{q}^{\alpha}_i \gamma_\mu q^{\alpha}_i,
\end{eqnarray}
where $Y$ is the hypercharge and the color indices are written explicitly as well as the flavor indices. Due to the mixing in the neutral boson sector, $u$ and $d$ quark or $e$ and $\nu_e$ have different charges. The electromagnetic interaction between $u$ quark and photon is different from that between $d$ quark and photon. Therefore, the final state is not necessary to be flavor singlet even though the initial state is flavor singlet because the $SU(2)_f$ symmetry is not the common symmetry of the strong and electroweak interactions. As a result, for example, for the $e^+e^- \rightarrow q\bar{q}$ process, $u\bar{u}$ and $d\bar{d}$ or $p\bar{p}$ and $n\bar{n}$ can be created with different possibilities resulting in flavor non-singlet state.

The electroweak interaction is color blind and it is the same for the quark with color red, green and blue. As well as the $SU(2)_L\times U(1)_Y$ invariance, the Lagrangian of the electroweak interaction is invariant under the global $SU(3)_c$ transformation of (\ref{eq:SU3}) with $\theta^a(x)$ being $x$ independent constants. In fact, the currents in Eq. (\ref{eq:currents}) can be expressed as $\bar{q}_i^\alpha \Gamma_{ij} q_j^\alpha$ in general, where $\Gamma$ includes flavor and $\gamma$ matrices. The currents are invariant under the global $SU(3)_c$ transformation as 
\begin{equation}\label{eq:current}
(\bar{q}')_i^{\beta} \Gamma_{ij} (q')_j^\beta = \bar{q}_i^{\gamma} {\rm exp}\left( -i\theta^a \frac{\lambda^a}{2}\right)_{\gamma\beta} \Gamma_{ij} {\rm exp}\left( i\theta^a \frac{\lambda^a}{2}\right)_{\beta\alpha} q_j^\alpha 
= \bar{q}_i^{\gamma} \Gamma_{ij} \delta_{\gamma \alpha}q_j^\alpha
=\bar{q}_i^\alpha \Gamma_{ij} q_j^\alpha,
\end{equation} 
where $\Gamma$ commutes with the color matrices $\lambda^a$.
For leptons and the gauge fields $W_\mu$, $Z_\mu$, $A_\mu$ in the electroweak interaction, they are unchanged under the color transformation since they are colorless without color index. Here we did not include the mixing of three generations in flavor space since it does not affect the $SU(3)_c$ invariance. One can see though the strong and electroweak interactions have different local symmetries, all the fundamental interactions ${\cal L}_I$ in the standard model are globally $SU(3)_c$ invariant as ${\cal L}'_I(x)={\cal L}_I(x)$.  
Therefore, if the initial state $|i \rangle $ (or the origin of the universe) is a color singlet, i.e., $S |i \rangle = |i \rangle $, the final state $|j \rangle = {\rm exp}(-i\int d^4 x{\cal L}_I(x)) |i \rangle $ should be also a color singlet because 
\begin{equation}
S |j \rangle = S {\rm exp}(-i\int d^4 x{\cal L}_I(x)) S^{-1}S |i \rangle = {\rm exp}(-i\int d^4 x{\cal L}'_I(x))|i \rangle =|j \rangle,
\end{equation} 
where $S$ is the $SU(3)_c$ color transformation. 

\begin{figure}[tb]
\begin{center}
\includegraphics[scale=0.85]{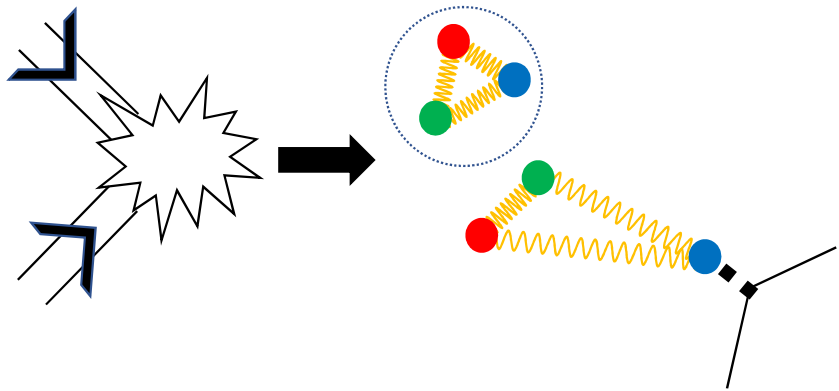}
\caption{Schematic diagram for producing and detecting the colored quark. The states on the left and right sides of the arrow are both color singlets. The separation of a single quark or detecting the color of the quark (the most right one) will fix its color and make the final state of the whole system color non-singlet.}
\end{center}
\end{figure}

A schematic diagram for producing and detecting the colored quark is plotted in Fig. 1. In any physical process, colored quarks and gluons can be produced from the interaction of the initial color singlet states. As explained above the final colored particles have to be combined into color singlet objects, such as baryons, mesons, tetraquarks, pentaquarks, di-baryons etc, because all the fundamental interactions are $SU(3)_c$ invariant. For example, the wave functions of the color parts of meson and baryon are $\delta_{ab}q_a\bar{q}_b$ and $\epsilon_{abc} q_a q_b q_c$, where $q_1=q_r$, $q_2=q_g$ and $q_3=q_b$. Though each quark or antiquark has the color, the color is not fixed and it runs over red, green and blue. If we separate the most right quark in Fig. 1 by any interactions of the standard model between this quark and experimental color singlet facility, the whole system including the facility should be color singlet before and after the separation. The separation of a single quark or the detection of the color of a single quark with any method (interaction) will fix the color and make the final state non-singlet (destroy the color singlet state) which is impossible due to the $SU(3)_c$ invariance of the fundamental interactions. This is comparable with the spin state. For example, if two electrons form a spin-zero state, the spin of each electron can be both $1/2$ and $-1/2$. The detection of the spin of one electron will make its spin fixed to be either $1/2$ or $-1/2$ and the spin-zero state will be destroyed. The difference is, for spin case, the spin of the electron can be measured because the total spin of whole system including the detecting facility is not required to be zero. For color case, we can not detect the color because the whole system has to be color singlet. 
Therefore, without the detailed information of the potential between colored objects, or no matter what kind of potential between colored objects, we can conclude quark or colored object can never be separated due to the common symmetry of the standard model.

\begin{figure}[tbph]
\begin{center}
\includegraphics[scale=1.0]{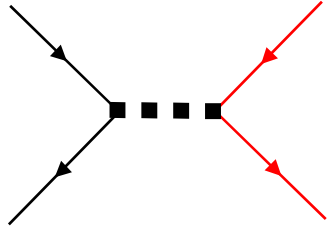}
\caption{Leading diagram for the interaction between electron and quark or between color singlet state and quark. The red lines are for quark fields, while solid black lines are for electrons or color singlet states. The dashed line is for the electromagnetic and strong interactions.}
\end{center}
\end{figure}

We should emphasis the wave function of the color part is somewhat different from the traditional wave function in quantum mechanics. In coordinate space, the wave function is defined as 
$\Psi(\vec{x})=\langle \vec{x}|\Psi \rangle $, where $|\Psi(\vec{x})|^2$ is the possibility of the particle located at $\vec{x}$. The wave function can be expanded in terms of momentum and energy eigenstates as 
\begin{equation}
\Psi(\vec{x})= \int \frac{d^3 \vec{p}} {(2\pi)^3} \psi(\vec{p}) e^{i\vec{p}\cdot \vec{x}} = \sum_n c_n \phi_n(\vec{x}),
\end{equation} 
where $\psi(\vec{p})=\langle \vec{p}|\Psi \rangle $ and $c_n= \langle n|\Psi \rangle $. $|\psi(\vec{p})|^2$ and $|c_n|^2$ are the possibilities to find the state with momentum $\vec{p}$ and energy $E_n$, respectively. For a given state $\Psi(\vec{x})$, we can measure the momentum or energy of the state with the interaction of the detecting facility and the state (particle). In other words, the state can be changed to be momentum or energy eigenstate with the corresponding possibility as long as the total momentum and energy of the whole system are conserved. However, the color wave function can not be changed by the measurement due to the $SU(3)_c$ invariance. For example, for a quark-antiquark state, the color singlet wave function is $\Psi_c= \frac{1}{\sqrt{3}} \delta_{\alpha\beta} \alpha \bar{\beta}$, where $\alpha,\beta = r, g, b$. Though direct calculation of the possibility $\langle r\bar{r}|\Psi_c \rangle^2$ is $\frac13$, this possibility of the state with red quark and anti-quark can only be measured or detected by the $SU(3)_c$ breaking interaction. Therefore, in the calculation of the cross section for such process, we should be very careful for the color indices because the measurement can not change the state from a color singlet to a color non-singlet. For example, for the $e^+e^- \rightarrow q\bar{q}$ case, the cross section $\sigma$ can be easily calculated by the tree diagram in Fig.~2 with the final state a color singlet. However, the cross section of $e^+e^- \rightarrow q_r\bar{q}_r$ is zero instead of $\frac13 \sigma$.

To be more clear, let's look at the electromagnetic interaction of quark and photon which can be written as 
\begin{equation}\label{eq:LQED}
{\cal L}_I^{\rm em} = e_q J^\mu (x) A_\mu (x) = e_q\bar{q}_\alpha(x) \gamma^\mu q_\beta(x) \delta_{\alpha\beta} A_\mu(x),
\end{equation}
where $J^\mu(x)$ is the color singlet current ($SU(3)_c$ invariant current). The color structure of the quark electromagnetic current in Fig.~2 can not be changed by any measurement. It can only be a color singlet with the sum of all the three colors. The cross section of $e^+e^- \rightarrow q_r\bar{q}_r$ is zero if there is no $SU(3)_c$ breaking interaction. The situation is the same if the $s$ channel process is changed to be the $t$ channel scattering case, i.e., there is no interaction between color singlet states and colored quark. From Eq.~(\ref{eq:LQED}), one can see the colorless photon can only be emitted from the color singlet current where the color structure can not be changed by any $SU(3)_c$ invariant interaction. Similar as for the electromagnetic interaction, there is no strong interaction by multi-gluon exchange between color singlet states and colored quark either, because the color structure of any color singlet quark current $J$ has to be $\bar{q}_\alpha(x) \Gamma q_\beta(x) \delta_{\alpha\beta}$, where $\Gamma$ represents the $\gamma$ matrices. $\alpha$ and $\beta$ are the colors of the initial and final quarks. The indices $\alpha$ and $\beta$ must run over the three colors to make the interacting vertex $SU(3)_c$ invariant. In general, for any colored states, such as a colored quark $q_\alpha$, a color-triplet diquark $\epsilon_{\alpha\beta\gamma} q_\beta q_\gamma$, a color-octet gluon ${\cal A}_\mu^a$, etc, they have no interaction with the color singlet states because the color indices $\alpha$ and $a$ of the colored states are fixed. Otherwise, color singlet states will turn into a pair of quarks, diquarks or gluons with specified colors. This can never happen within the standard model which is globally $SU(3)_c$ invariant.
 
To interact with color singlet states, colored states have to be combined with the other colored particles to form into color singlet states. Besides the electroweak and strong interactions between singlet states, these interactions also exist within the color singlet states. In both cases, the color indices of quark and gluon in every vertex are summed up to guarantee that the interacting vertex is $SU(3)_c$ invariant. For example, in a baryon formed by a colored quark ($q_\alpha$) and a colored diquark ($\epsilon_{\alpha\beta\gamma} q_\beta q_\gamma$), the color factors of the interactions by one-photon and one-gluon exchanges are $\epsilon_{\alpha\beta\gamma}\epsilon_{\alpha\beta'\gamma'}\delta_{\beta\beta'}\delta_{\gamma\gamma'}=6$
and $\epsilon_{\alpha\beta\gamma}\epsilon_{\alpha\beta'\gamma'T^a_{\beta\beta'}T^a_{\gamma\gamma'}}=-16$, respectively. For gluons, to interact with color singlet states, they have to form into glueballs, hybrids, or other color singlet states so that the colors of the gluons become hidden indices. A color-octet gluon with color $a$ has no interaction with mesons and baryons.

Therefore, colored objects are reasonable candidates for dark matter. They have no interaction with the color singlet universe in the standard model. A question may rise. How is colored matter created from the big bang if the origin of the universe is $SU(3)_c$ invariant?
The question can be easily answered if we assume quarks with different colors have different sizes. For non-point particles, the nonlocal interaction was proposed for quantum electrodynamics as well as for the effective field theory \cite{He,Salamu,Yang,He2}. The virtue of the nonlocal formulation is that it allows the use of a 4-dimensional regulator while preserving the gauge and Lorentz symmetries.
Higgs mechanism is not affected except the couplings between Higgs particle and fermions are nonlocal. The mass terms generated from the nonzero expectation value of Higgs field in vacuum are not changed because the expectation value is spacetime independent.

Let's take the nonlocal QED as an example and the nonlocal realization for QCD is straightforward. The nonlocal $U(1)$ gauge invariant Lagrangian for quark and photon is written as \cite{He2}
\begin{equation}
\mathcal{L}_{\rm QED}^{\rm nl} = \bar{q}_\alpha(x)(i\slashed{\partial}-m)q_\alpha(x)-\frac14F_{\mu\nu}(x)F^{\mu\nu}(x)
-e_q\int\!\,d^4 a \bar {q}_\alpha(x)\gamma^\mu q_\alpha(x) A_\mu(x+a)F(a,\alpha),
\end{equation}
where the quark field $q_\alpha(x)$ is located at $x$ and the photon field $A_\mu(x+a)$ is located at
$x+a$. $F(a,\alpha)$ is the correlation function normalized as $\int d^4 a F(a,\alpha)=1$. If it is chosen to be a $\delta$ function, 
the nonlocal Lagrangian will be changed back to the local one. The above nonlocal Lagrangian is invariant
under the following gauge transformation
\begin{equation}
\psi(x)\rightarrow\,e^{i\theta(x)}\psi(x),~~~~~~~~A_\mu\rightarrow\,A_\mu-\frac{1}{e_q}\partial_\mu\,\theta^\prime(x),
\end{equation}
where $\theta(x)=\int\,da\theta^\prime(x+a)F(a,\alpha)$. If $F(a,r)=F(a,g)=F(a,b)$, the nonlocal Lagrangian is globally $SU(3)_c$ invariant as the local QED. If the correlation functions are different for quarks with different colors, i.e., the couplings for photon and quarks with different colors are not identical, the nonlocal QED breaks $SU(3)_c$ symmetry. 
With the correlation function, the quark-photon vertex is momentum dependent as $e_q\gamma_\mu\,\tilde F(k,\alpha)$, where $\tilde F(k,\alpha)$ is Fourier transformation of the correlation function $F(a,\alpha)$ and $k$ is the photon momentum. Let's take the dipole function as an example \cite {He2}
\begin{equation}
\tilde F(k,\alpha) = \frac{\Lambda_\alpha^4}{(k^2-\Lambda_\alpha^2)^2},
\end{equation}
where $\Lambda_\alpha$ is the color-dependent parameter.
Since quark could be very small, $\Lambda_\alpha$ is very large and the size effect will only be visible at very large $k^2$. If $k^2$ is not large enough, $\tilde F(k,\alpha) \approx 1$ and $SU(3)_c$ breaking is invisible. The momentum (energy) transfer could be large enough for big bang so that the quark pairs with color red, green and blue are created with different numbers. These particles can form color singlet and color non-singlet states. When the universe cools down, the color singlet and color non-singlet states will be decoupled from each other and the color non-singlet objects turn into dark matter with nearly no interaction with the color singlet universe except gravity which may not be described in the framework of quantum theory. With the nonlocal interaction, on the one hand, color non-singlet objects can be created with extremely large momentum (energy) transfer from the big bang. On the other hand, these colored objects composed by the elemental particles in the standard model will turn into dark matter in the universe. 

With the simple assumption, the colored objects can be the candidate of dark matter. Currently, 
in addition to the weakly interacting massive particles, many extensions of the standard model have been proposed for studying the candidates of dark matter. The typical approach is to assume a hidden sector of new elementary particles with a weak connection to the visible particles by a mediator known as portal. There are some well-known portals, such as the vector or gauge boson portal \cite{Holdom}, the Higgs portal \cite{Patt}, the neutrino portal \cite{Macias,Blennow,Iwamoto} and the axion portal \cite{Nomura,Semertzidis}. Many collaborations around the world are building exquisitely sensitive detectors to search for dark matter particles. The search for new physics in low mass and weak coupling range is known as the low-energy high-precision frontier. During the last few decades, substantial experimental efforts have been made to search for the axion. Meanwhile, the direct and indirect detection searches for dark matter through neutrino portal are also carried out.
Compared with the other candidates of dark matter, the possibility of producing and detecting colored dark matter could be greatly increased with the increasing collision energy of the experiments.

We have pointed colored particles are created and form into color singlet and non-singlet states after the big bang. For any color, quark and anti-quark are created in pair and the total baryon number is conserved to be 0. We know the baryon number of the color singlet universe is positive and as a result, the baryon number of the color non-singlet objects, the dark matter, has to be negative, which is the ``disappeared" anti-matter. This naturally explains why there is large asymmetry of matter and anti-matter in the universe which are created with the same amount from the big bang. Anti-matter does not disappear and it exists as dark matter in the universe. Though the baryon number of colored dark matter is the same as that of antibaryons, they can not produce the color singlet antibaryons by any interactions within the standard model. This process is forbidden due to the $SU(3)_c$ invariance of the fundamental interactions. Colored and color singlet objects can only turn into each other by the $SU(3)_c$ breaking interactions.

The above theoretical conclusions can be confirmed by the experiments. Color non-singlet dark matter could be produced in the laboratory with the high energy collisions of lepton and anti-lepton or hadron and anti-hadron, where the colored fireball is created. It can be tested by measuring the energy loss and baryon number change of the final state. These kinds of experiments will definitely shed light on the knowledge of color confinement, dark matter and the missing of anti-matter in the universe.

In summary, we have shown a clear and simple proof for color confinement based on the common symmetry of the standard model and this long-standing problem is solved convincingly. The interesting realization is that we need pay attention to the other fundamental interactions rather than focus only on the strong interaction, though apparently, color confinement is originated from strong interaction. Color confinement is the direct consequence of the common symmetry of all the fundamental interactions. The linear potential or large coupling constant is not crucial for color confinement. With the nonlocal Lagrangian based on the simple assumption that quarks with different colors have different sizes, color non-singlet objects can be produced from the big bang. They are reasonable candidates for dark matter which has nearly no interaction with the color singlet universe except gravity. Anti-matter does not disappear and it exists as dark matter in colored states. The dominance of matter over anti-matter in the universe can be easily understood. We predict dark matter could be produced in the laboratory and this could be tested by measuring the energy loss and baryon number change in the extremely high energy collisions of particles and anti-particles.

\section*{Acknowledgments}
This work is supported by the National Natural Science Foundation of China under Grant No.~11975241.


\begin{thebibliography}{10}
\bibitem{Hooft} Gerard 't Hooft and M. J. G. Veltman, Nucl. Phys. B 44 (1972) 189.
\bibitem{Gross} D. J. Gross and F. Wilczek, Phys. Rev. Lett. 30 (1973) 1343.
\bibitem{Politzer} H. D. Politzer, Phys. Rev. Lett. 30 (1973) 1346.
\bibitem{Konishi} K. Konishi and L. Spanu, Int. J. Mod. Phys. A 18 (2003) 249.
\bibitem{Takahashi} T. T. Takahashi, H. Matsufuru, Y. Nemoto and H. Suganuma, Phys. Rev. Lett. 86 (2001) 18.
\bibitem{Nakamura} A. Nakamura and T. Saito, Prog. Theor. Phys. 115 (2006) 189.
\bibitem{Kharzeev} D. E. Kharzeev and E. M. Levin, Phys. Rev. Lett. 114 (2015) 242001.
\bibitem{Kuvshinov} V. I. Kuvshinov and E. G. Bagashov, Theor. Math. Phys. 184 (2015) 1304.
\bibitem{Jungman} G. Jungman, M. Kamionkowski and K. Griest, Phys. Rep. 267 (1996) 195.
\bibitem{Byrne} M. Byrne, C. Kolda and P. Regan, Phys. Rev. D 66 (2002) 075007.
\bibitem{Peccei} R. D. Peccei and H. R. Quinn, Phys. Rev. Lett. 38 (1977) 1440
\bibitem{Abe} N. Abe, T. Moroi and M. Yamaguchi, J. High Energy Phys. 0201 (2002) 010.
\bibitem{Dodelson} S. Dodelson and L. M. Widrow, Phys. Rev. Lett. 72 (1994) 17. 
\bibitem{Carr}B. Carr, F. Kuhnel and M. Sandstad, Phys. Rev. D 94 (2016) 083504.
\bibitem{Arun} K. Arun, S.B. Gudennavar and C. Sivaram, Adv. Space Res. 60 (2017) 166.
\bibitem{Bernabei} R. Bernabei et al.[DAMA and LIBRA Collaborations], Eur. Phys. J. C 73 (2013) 2648.
\bibitem{Akeribet} D. S. Akerib et al (LUX Collaboration), Phys. Rev. Lett. 112 (2014) 091303.
\bibitem{Tan} A. Tan et al (PandaX-II Collaboration), Phys. Rev. Lett. 117 (2016) 121303.
\bibitem{Amole} C. Amole et al (PICO Collaboration), Phys. Rev. Lett. 118 (2017) 251301.
\bibitem{Agnese} R. Agnese et al (SuperCDMS Collaboration), Phys. Rev. Lett. 120, 061802 (2018)
\bibitem{Abdelhameed} A. H. Abdelhameed et al (CRESST Collaboration), Phys.Rev.D 100 (2019) 102002.
\bibitem{Langacher} P. Langacker, Adv. Ser. Direct. High Energy Phys. 3 (1989) 552.
\bibitem{Bernreuther} W. Bernreuther, Lect. Notes Phys. 591 (2002) 237.
\bibitem{Weinberg} S. Weinberg, Phys. Rev. Lett. 19 (1967) 1264.
\bibitem{Glashow} S. L. Glashow, Nucl. Phys. 22 (1961) 579.
\bibitem{Salam} A. Salam, Conf. Proc. C 680519 (1968) 367.
\bibitem{He} Fangcheng He and P. Wang, Phys. Rev. D 97 (2018) 036007.
\bibitem{Salamu} Y. Salamu, C. R. Ji, W. Melnitchouk, A. W. Thomas and P. Wang, Phys. Rev. D 99 (2019) 014041.
\bibitem{Yang} Mingyang Yang and P. Wang, Phys. Rev. D 102 (2020) 056024.
\bibitem{He2} Fangcheng He and P. Wang, Eur. Phys. J. Plus 135 (2020) 156.
\bibitem{Holdom} B. Holdom, Phys. Lett. B 166(1986) 196.
\bibitem{Patt} B. Patt and F. Wilczek, hep-ph/0605188.
\bibitem{Macias} V. G. Macias and J. Wudka, JHEP 07 (2015) 161.
\bibitem{Blennow}  M. Blennow $et$ $al$, Eur. Phys. J. C 79 (2019) 555.
\bibitem{Iwamoto} S. Iwamoto, K. Seller and Z. Trocsanyi, arXiv:2104.11248.
\bibitem{Nomura} Y. Nomura and J. Thaler, Phys. Rev. D 79 (2009) 075008.
\bibitem{Semertzidis} Y. K. Semertzidis and S. Youn, arXiv:2104.14831.

\end{thebibliography}
\end{document}